%% file: sample-sigconf.tex
\documentclass[sigconf]{acmart}
\usepackage{booktabs} 
\usepackage{listings}
\usepackage{color}
\usepackage{algorithm}
\usepackage{algorithmic,algorithm}
\usepackage{multirow}
\usepackage{subfig}

\usepackage{footnote}
\makesavenoteenv{tabular}
\makesavenoteenv{table}

\definecolor{mygreen}{rgb}{0,0.6,0}
\definecolor{mygray}{rgb}{0.5,0.5,0.5}
\definecolor{mymauve}{rgb}{0.58,0,0.82}
 \setcopyright{none}

\acmDOI{}

\acmConference[]{}{}{}
\acmYear{2019}
\copyrightyear{2019}

\begin{document}
\title{Accelerating Distributed Deep Learning Training with \\ Gradient Compression}

 \author{Jiarui Fang, Haohuan Fu and Guangwen Yang}
\affiliation{%
  \institution{$\textit{Tsinghua University}$}
  \institution{$\texttt{fjr14@mails.tsinghua.edu.cn}$}
   \institution{$\texttt{\{haohuan,ygw\}@tsinghua.edu.cn}$}
}

\author{Cho-Jui Hsieh}
 \authornote{This work was done when Jiarui Fang was a visiting scholar at Cho-Jui Hsieh's group.}
\affiliation{%
  \institution{$\textit{University of California, Los Angeles}$}
  \institution{$\texttt{chohsieh@cs.ucla.edu}$}
 }







\begin{abstract}
Data parallelism has become a dominant method to scale Deep Neural Network (DNN) training across multiple nodes. 
Since synchronizing a large number of gradients of the local model can be a bottleneck for large-scale distributed training,
compressing communication data has gained widespread attention recently. 
Among several recent proposed compression algorithms, 
Residual Gradient Compression (RGC) is one of the most successful approaches---it can significantly compress the transmitting message size (0.1\% of the gradient size) of each node and still achieve correct accuracy and the same convergence speed.
However, the literature on compressing deep networks focuses almost exclusively on achieving good theoretical compression rate, while the efficiency of RGC in real distributed implementation has been less investigated.
In this paper, we develop an RGC-based system that is able to reduce the end-to-end training time on real-world multi-GPU systems.
Our proposed design called RedSync, 
which introduces a set of optimizations to reduce communication bandwidth requirement while introducing limited overhead.
We evaluate the performance of RedSync on two different multiple GPU platforms, including 128 GPUs of a supercomputer and an 8-GPU server.
Our test cases include image classification tasks on Cifar10 and ImageNet, and language modeling tasks on Penn Treebank and Wiki2 datasets.
For DNNs featured with high communication to computation ratio, which have long been considered with poor scalability, RedSync brings significant performance improvements.
\end{abstract}

%
%

\keywords{Deep Learning System, Data Parallel, Gradient Compression}

\maketitle

\input{samplebody-conf}

\bibliographystyle{ACM-Reference-Format}
\bibliography{fullbib.bib}

\end{document}

%% file: samplebody-conf.tex
\section{Introduction}
\label{}
\begin{figure}[ht!]
\centering
\includegraphics[width=0.45\textwidth]{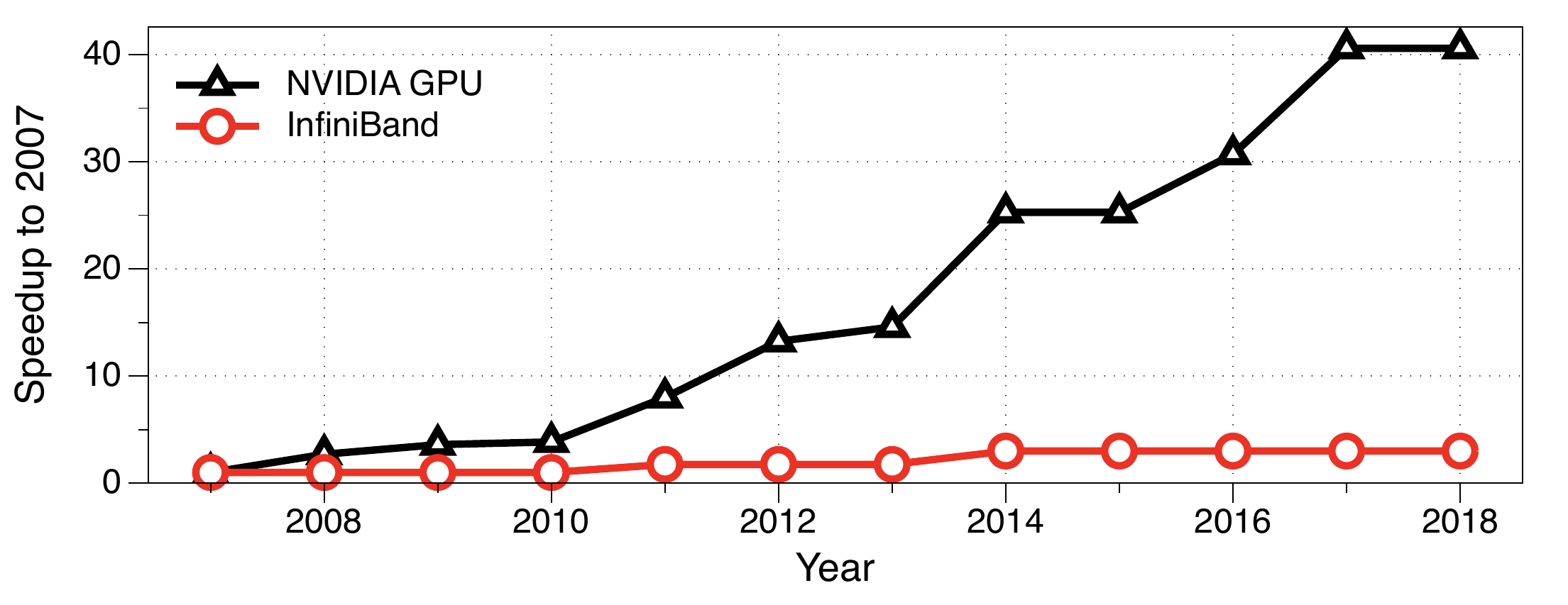}
\caption{Evolution Speed of Infiniband and GPU.}
\label{fig:IBvsGPU}
\end{figure}

For training large-scale deep neural networks (DNNs) on multiple computing nodes, data parallelism has emerged as the most popular choice due to its simplicity and effectiveness \cite{dean2012large, recht2011hogwild}. 
However, the limited communication bandwidth of the interconnected network has become the bottleneck limiting data parallel performance. 
First, models of DNNs, which already contain tens to hundreds of layers and totaling 10-20 million parameters today, continue to grow bigger~\cite{real2018regularized}.
Therefore, the requirement of fast synchronizing model parameter updates among all computing nodes poses a greater challenge.
Second, the development of DNN-customized training accelerators has shifted the bottleneck of training from computing hardware towards communication across nodes.
Meanwhile, the evolution of the interconnected network is not as fast as computing hardware, and this trend still continues.
As shown in Figure\ref{fig:IBvsGPU}, in the past decade, the most powerful GPU has been over 40 times faster, comparing the latest Tesla V100 GPU with G8800 GPU.
However, the network bandwidth of switches in clusters is only 5 times faster, comparing InfiniBand HDR with InfiniBand QDR.
Third, high-quality network fabric like InfiniBand is too expensive to be available for every data center.
Publicly available cloud computing resources are still connected by the low-level network.
For example, Amazon EC2 instances now provide a maximum bandwidth of 25 Gbps, far less than 96 Gbps of InfiniBand EDR 4-Link speed.

Many recent studies focused on reducing the communication cost between nodes by reducing the size of the gradients to be transmitted.
One line of work \cite{seide20141, wen2017terngrad} propose to quantize the gradients to low-precision values.
Considering compression ratio (ratio of compressed gradients size to their original size) achieved by quantization is limited, another line of research orthogonal to quantization is to sparsify communication gradients and restrict weight-updates to a small subset of parameters.
Residual Gradient Compression (RGC) method \cite{strom2015scalable, aji2017sparse, chen2018adacomp, lin2017deep, sattler2018sparse} is currently the most promising sparsification method to achieve good compression ratio while ensuring no loss of training accuracy.
It transmits only a small subset of gradients and maintains the remaining gradients locally as residuals to be added to gradients of the next iteration.
The first RGC implementation is proposed by Strom\cite{strom2015scalable} and uses a threshold-based method to only send gradients larger than a predefined constant threshold for fully-connected layers. 
Considering a predefined threshold is hard to be chosen appropriately, Aji et. al.\cite{aji2017sparse} improve the robustness of RGC by selecting top 1\% gradients to communicate according to their magnitude.
Because these two implementations are tuned for some specific network structures, applying them to other DNNs will lead to accuracy loss as indicated in AdaComp\cite{chen2018adacomp}.
Based on their work, after introducing some key modifications, the latest RGC variant called DGC\cite{ lin2017deep} is able to achieve a 0.1\% compression ratio on local gradients while ensuring almost no loss of model accuracy on a variety of  DNN structures.

Despite good model accuracy achieved with simulation experiments, no recent studies have discussed the potential performance gain after integrating the latest RGC methods to a real distributed training system,
especially to the multi-GPU systems equipped with high-quality network infrastructures.
The challenges of applying RGC to distributed GPU systems come from two aspects.
First, there is no efficient compression algorithm of RGC method designed for massive parallel processors such as GPU.
According to our experimental results of Section \ref{sec:select}, selecting top-0.1\% elements with the state-of-the-art GPU-based top-$k$ algorithm are so expensive that the overhead of compression is much higher than the benefits of network bandwidth reduction.
Second, the synchronization scheme of sparse data structures generated by the RGC method has not been well studied. It is not easy to be supported with existing efficient communication libraries, such as Message Passing Interface (MPI), which are designed for dense data structures.

To increase scalability and efficiency of DNN training, we propose a systematic distributed design called RedSync (short for \textbf{Red}uction of \textbf{Sync}hronization Bandwidth), which combines RGC-based sparsification and quantization techniques together to compress transmitting gradient size of each node to its 0.1\%.
Our contributions are listed as follows:
\begin{itemize}
\item
In terms of the algorithm level design, we propose a quantization technique called Alternating Signs Quantization (ASQ)  to further compress the size of transmitting data sparsified by RGC to its half.
In addition, we adapt the-state-of-the-art algorithmic improvements of RGC to a distributed situation.
Applying the proposed algorithmic improvements, RGC+ASQ improves the efficiency of RGC-only in tasks of training a set of state-of-the-art DNN models with no accuracy loss.
\item
In terms of the system level design, we remove two main obstacles for efficient RGC deployment.
A set of parallel-friendly top-0.1\% selection algorithms are proposed to support the sparsification process.
They are orders of magnitude faster than the state-of-the-art  top-$k$ selection method on GPU.
Considering the distribution characteristics of communication data, we apply Allgather operation using MPI for a sparse synchronization scheme.
A cost model is derived to analyze both communication cost and calculation overhead.
Based on it, we pointed out potential performance gain and the bottleneck of current RGC method.
\item
This is the first work, as far as we know, to evaluate the performance of RGC method on supercomputer scale.
On 128 GPUs, RedSync provides significant performance improvements for communication-intensive networks, like VGG, AlexNet, and some LSTMs.
\end{itemize}




\section{Design of RedSync}

\begin{algorithm}[ht!]
  \small
     \caption{\small{RedSync Workflow}}
        \label{algo:rgcframework}
        \begin{algorithmic}
         \renewcommand{\algorithmicrequire}{\textbf{Input:}}
         \renewcommand{\algorithmicensure}{\textbf{Output:}}
         \REQUIRE node id $k$; the number of node $N$
         \REQUIRE training dataset $\chi$; mini batch size $b$ per node
        	\REQUIRE initial model $\textbf{w} = {\textbf{w}[0], ... , \textbf{w}[\#layer]}$; compression ratio $D$
        	\STATE $V^k \gets 0$
         \FOR {$t = 0,1, ... max\_iter$}
                	\STATE sample $b$ elements as $\chi_k^t$ 
		\STATE  $G^k$ $\gets \nabla$ $f$($\chi_k^t$ ; \textbf{w}) : forward and backward propagation
        		\FOR {$j = \#layer, \#layer-1, ..., 0$}
        			\STATE $V_j^k += G_j^k $
        			\STATE {Masks} $\gets$ $\tt{select}$ $(V_j^k, D)$
        			\STATE $G_j^k$ $\gets$ $\tt{Sparse\-Allreduce}$($\tt{compress}$($\tt{quantize}$($V_j^k$ $\odot$ Masks)))
        			\STATE $V_j^k \gets V_j^k \odot$  (1 - Masks)
        		\ENDFOR
        		\STATE \textbf{w} $\gets$ SGD(\textbf{w}, $\tt{decompress}$($G^k$))
                \ENDFOR
        \end{algorithmic}
\end{algorithm}

\begin{figure}[ht!]
\centering
\includegraphics[width=0.45\textwidth]{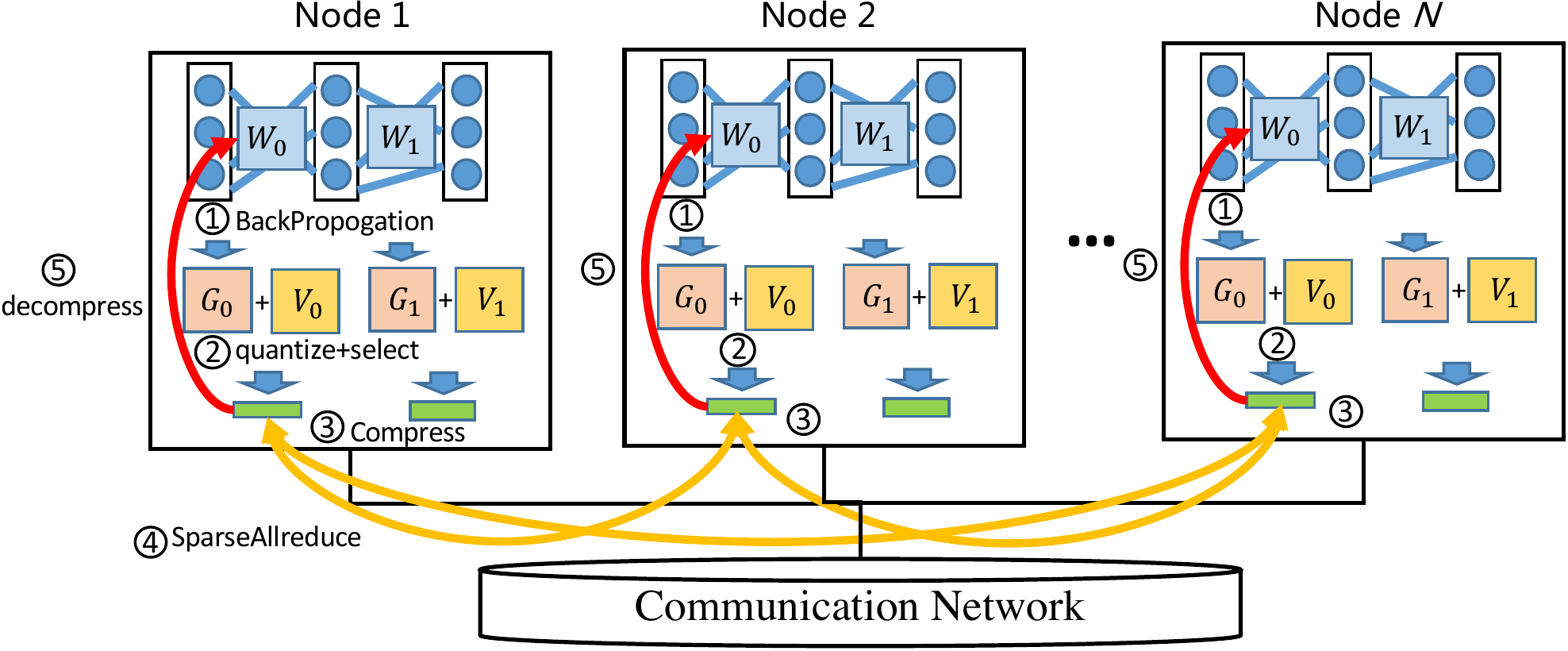}
\caption{Overview of RedSync Design.}
\label{fig:RedSync}
\end{figure}

Algorithm \ref{algo:rgcframework} presents the workflow used in RedSync.
We denote a DNN model as $f(\textbf{w})$, where $\textbf{w}$ is the vector of parameters.
We assume a system has $N$ workers. Each worker, say the $k$-th worker, holds a local dataset $\chi_k^t$ at iteration $t$ with size $b$ and a local copy of the global weight $\textbf{w}$.
Synchronous SGD method is adopted in RedSync.
At each iteration, node $k$ computes the gradient  $G^k$ using local data and we represent $G_j^k$ as gradients of layer $j$.
Each node also maintains a residual $V^k$, which is initialized as $0$ and used to accumulate untransmitted gradient from previous iterations. 
After added with latest gradient, a subset of residuals is selected as the \textit{communication-set}.
The $\tt{select}$ operation in Algorithm \ref{algo:rgcframework} chooses important elements as communication-set based on magnitude. 
Masks is a 0/1 matrix, in which 1 indicating that the element at the corresponding position is selected as communication-set.
After $\tt{quantize}$ operation, the quantized communication-set is compressed into sparse data structures for communication.
Those selected elements are synchronized among all the nodes using $\tt{Sparse\-Allreduce}$ operations.
Synchronous SGD implemented with Allreduce, rather than Parameter Server\cite{li2014scaling}, has been widely adopted in state-of-the-art large-scale DNN training tasks\cite{goyal2017accurate}\cite{you2017imagenet} on HPC platforms. 
Remaining elements outside the communication-set are assigned as new residuals of the next iteration.
Figure \ref{fig:RedSync} presents the design overview of RedSync according to the Algorithm \ref{algo:rgcframework}.
In the following, we details our contribution in design of $\tt{select}$, $\tt{quantize}$, $\tt{Sparse\-Allreduce}$ and $\tt{decompress}$ operations to make this workflow work efficient in practice.

\subsection{Parallel-Friendly Residual Sparsification}
\label{sec:select}
\begin{figure}[ht!]
\centering
\includegraphics[width=0.5\textwidth]{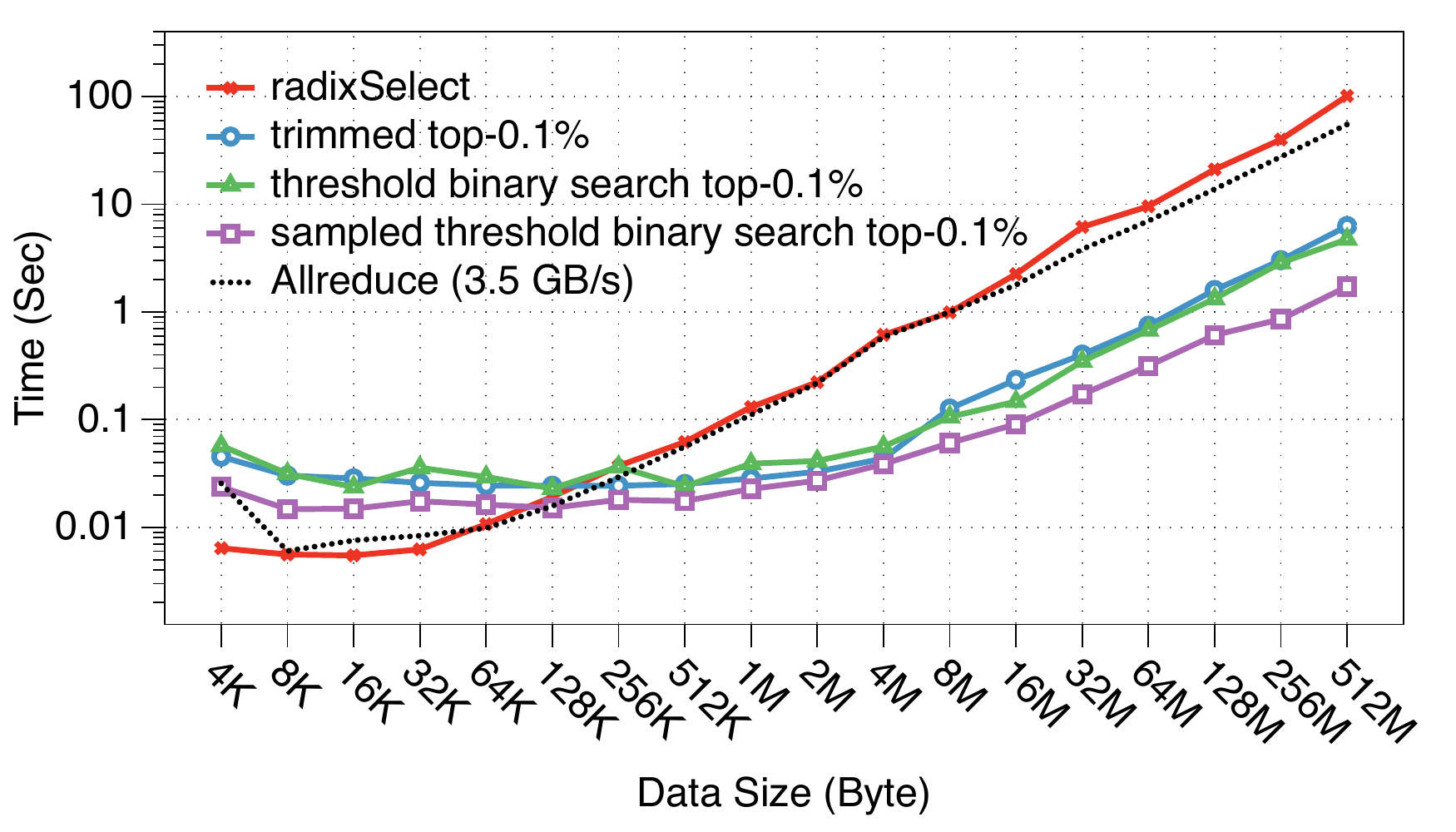}
\caption{
       Performance of four communication-set selection algorithms vs Data sizes. 
}
\label{fig:select_time}
\end{figure}

The efficiency of communication-set selection operation to sparsify the gradients is critical to the overall performance of the system.
There is still a lack of efficient implementation methods on parallel architectures such as GPUs, which is a major obstacle to the deployment of RGC methods.
Recent RGC works \cite{lin2017deep, sattler2018sparse} suggest selecting top $0.1\%$ elements from residuals of each layer as the communication-set.
It is well-known that, by applying \textit{Quickselect} algorithm \cite{hoare1961find}, the time complexity of a top-$k$ selection on a list of $n$  elements using a single-core CPU is $O(n)$.
However, the top-$0.1\%$ selection is not trivial to be implemented on massively parallel architectures, such as GPUs.
One of the most efficient top-$k$ selection methods designed for GPU can be implemented based on \textit{radixSelect} algorithm \cite{alabi2012fast}, 
which determines each bit of the $k$-th largest element by scan and scatter. 
The \textit{scan} \cite{sengupta2007scan} and scatter operations are not suitable for parallel processing and are extremely time-consuming.
As shown in Figure \ref{fig:select_time}, the computation time for top-0.1\% with radixSelect on a Titan X GPU sometimes is even slightly higher than the time for synchronizing these parameters through a 28 Gbps network.
To overcome the problem of slow radixSelect operation on GPU,
we propose two communication-set selection algorithms called \textit{trimmed top-$k$ selection} and \textit{threshold binary search selection}, which are more efficient on GPUs.

\textbf{Trimmed top-0.1\% selection.} 
We notice that  the distribution of residuals is usually similar to a normal distribution, we can use statistical features to remove most of the smaller elements and apply radixSelect operation on a relatively small subset.
As shown in Algorithm \ref{algo:topk}, we first calculate the mean and maximum of residuals' absolute values of this layer.
A relative large threshold value is chosen according to mean and maximum value, for example, $0.8\times(max - mean) + mean$.
Operation $\tt{count\_nonzero}$ gets the number of elements whose absolute values are greater than the threshold.
If the number is smaller than $k$ (the number of top-0.1\% elements ), we dynamically decrease the threshold until we find the number of parameters whose absolute value above the threshold is larger than $k$.
Then we trim all elements that are less than the threshold and perform a top-$k$ selection operation using radixSelect on the remaining elements.
Operation $\tt{mean}, \tt{max}$ and $\tt{count\_nonzero}$ can all be efficiently implemented with a single reduction operation.
$\tt{nonzero\_indices}$ is a typical \textit{stream compaction} problem, which uses just one scan operation as its backbone \cite{sengupta2006work}.

\textbf{Threshold binary search selection.} 
For a network layer with a large number of parameters,
even using radixSelect operation on a small number of gradient elements is still a very time consuming operation.
In order to completely avoid using radixSelect operation on GPU, we propose a method to select approximate top-0.1\% elements as communication-set.
Instead of identifying the $k$th (top 0.1\%th) largest element, we search for a threshold to make it between the $k$th to $2k$th largest element, and then select elements larger than the threshold as communication-set.
In this case, at least 0.1\% largest elements are included in the communication-set, so the convergence rate of the algorithm will not be affected.
As shown in Algorithm \ref{algo:thd}, we use a binary search algorithm to find such a threshold.
To avoid over-searching, the algorithm will automatically terminate when the difference between the left and right borders is less than the small value $\epsilon$.

For layers with large sizes, such as the first fully-connected layer in VGG16 and softmax layer in LSTM,
the time for $\tt{count\_nonzero}$ operation is still not negligible.
We further improve the efficiency of the selection algorithm by reducing the number of $\tt{count\_nonzero}$ operations.
We recommend that, after a threshold binary search for this layer, the threshold element can be reused in the next few iterations.
The interval of search is empirically set to 5, and the selection algorithm introduces only one $\tt{nonzero\_count}$ overhead on average.
Such a method is called sampled Threshold binary search selection.

\begin{algorithm}[ht!]
    \caption{Trimmed Top-0.1\%  Selection}
    \label{algo:topk}
    \begin{algorithmic}[1]
     \renewcommand{\algorithmicrequire}{\textbf{Input:}}
     \renewcommand{\algorithmicensure}{\textbf{Output:}}
     	\REQUIRE tensor to be compressed $X$
	\REQUIRE number of elements remained $k$
	\ENSURE $<indices, values>$
	\STATE $mean$ $\gets$ $\tt{mean}$(abs($X$)); $max$ $\gets$ $\tt{max}$(abs($X$))
	\STATE $\epsilon \gets 0.2$; $ratio \gets (1-\epsilon)$
	\STATE $nnz$ = $\tt{count\_nonzero}$(abs($X$) $>$ $threshold$)
	\WHILE{$nnz < k$}
		\STATE $threshold \gets mean + ratio \times (max - mean)$
		\STATE $nnz$ = $\tt{count\_nonzero}$(abs($X$) $>$ $threshold$)
		\STATE $ratio = ratio-\epsilon$
	\ENDWHILE
	\STATE $indices$ $\gets$ $\tt{nonzero\_indices}$(abs($X$) $>$ $threshold$))
	\STATE $values \gets X[indice]$
	\end{algorithmic}
\end{algorithm}

\begin{algorithm}[ht!]
    \caption{Top-0.1\% using Threshold Binary Search Selection}
    \label{algo:thd}
    \begin{algorithmic}[1]
     \renewcommand{\algorithmicrequire}{\textbf{Input:}}
     \renewcommand{\algorithmicensure}{\textbf{Output:}}
     	\REQUIRE tensor to be compressed $X$
	\REQUIRE number of elements remained $k$
	\REQUIRE Termination condition parameter $\epsilon$
	\ENSURE $<indices, values>$
	\STATE $mean$ $\gets$ $\tt{mean}$(abs($X$)); $max$ $\gets$ $\tt{max}$(abs($X$))
	\STATE $l \gets 0.0$; $r \gets 1.0$; $threshold = 0.0$
	\WHILE{$r - l > \epsilon$}
		\STATE $ratio = l + (r - l) / 2$
		\STATE $threshold \gets mean + ratio \times (max - mean)$
		\STATE $nnz$ = $\tt{count\_nonzero}$(abs($X$) $>$ $threshold$)
		\IF{$nnz > k$ and $2k > nnz$}
			\STATE break
		\ELSIF {$nnz < k/2$}
			\STATE $r = ratio$
		\ELSE
			\STATE $l = ratio$
		\ENDIF 
	\ENDWHILE
	\STATE $indices \gets $ $\tt{nonzero\_indices}$(abs($X$) $>$ $threshold$))
	\STATE $values \gets X[indices]$
	\end{algorithmic}
\end{algorithm}

In Figure \ref{fig:select_time}, we compared the time cost of different selection approaches applied to data of different sizes.
Test data is generated from a standard uniform distribution. 
$\tt{Allreduce}$ indicates the time cost to synchronize messages using the Allreduce operation, with a peak network bandwidth of 28 Gbps.
Performance is measured as total time cost for 100 times independent operations.
Compared with directly performing radixSelect, both proposed methods significantly reduce the selection operation time for data of large size.
For top-0.1\% selection on 64MB elements, 
trimmed top-0.1\% and sampled threshold binary search selection are 38.13 and 16.17 $\times$ faster than radixSelect.
In practice, a hybrid compression strategy is adopted: 
For smaller parameter sets such as biases and batch norm layers, we do not compress residuals or directly use radixSelect to select top-0.1\% significant elements.
Trimmed top-0.1\% selection is suitable for parameters of middle size layers, like convolutional layers, because it can ensure the compression ratio to be exactly 0.1\% and introduce no extra communication bandwidth requirements.
Threshold binary search based selection is suitable for large size layers, like hidden layers and softmax layers in LSTMs, for which the compression cost is more critical to be optimized than the communication cost.

\subsection{Quantization of Sparsified Residuals}
The sparsified communication-set to be transmitted includes $k$ index elements and $k$ value elements.
As indicated in Strom\cite{strom2015scalable}, by setting all value elements of the same sign to their average,
the communication bandwidth requirement of value elements can be eliminated by transmitting only one average element instead of $k$ elements.
As there are both positive and negative elements in the communication-set,
an extra data structure is used to record the sign of each element and extra efforts are required for quantization and de-quantization.

To eliminate this overhead, we design a quantization approach called \textit{Alternating Signs Quantization (ASQ)} to further reduce 1/2 of the bandwidth requirement.
In two adjacent training iterations, ASQ alternately quantizes the maximum 0.1\% elements and the minimum 0.1\% elements as communication-set instead of quantifying the maximum 0.1\% elements with the largest absolute value.
In other words, if we select the largest $k$ elements (all positive numbers) of this layer as the communication-set at current iteration, we will choose smallest $k$ elements (all negative numbers) as the communication-set for the next iteration.
Since top-0.1\% and bottom-0.1\% elements are all of the same sign, no communication bandwidth requires to transmit extra sign information.

As shown in experimental results, the quantized RedSync using ASQ approach is able to guarantee no accuracy loss.
As elements of residuals are designed to be delayed updated hundreds of steps, updating the elements of the same sign one more step later does not have signification impact on the direction of the gradient update.

ASQ introduces no quantization overhead and can be implemented efficiently by slightly modifying our parallel-friendly top-0.1\% approaches.
Compared with Strom's quantization method, ASQ approach is more memory-efficient and overhead-reduced.
Their method uses an extra bitmap to record the sign of each value element, which is also required to be transmitted.
Instead of once scan of entire data as ASQ, they have to separate positive and negative elements in the communication-set, and quantize them individually.
In addition, it is worth noting that \textit{sampled threshold binary search selection} cannot be used with quantization.
We also do not quantify the output layer of the DNN, in order to distinguish the correct classification information.

\subsection{Sparse Synchronization and Decompression}
\label{sec:sparseAllreduceComm}
Synchronization of dense gradient structures in traditional distributed DNN systems can be simply implemented with an Allreduce operation, which has been well-studied on multiple-GPU systems\cite{awan2017s}.
However, the design of a sparse Allreduce in a distributed setting is not as simple because each worker may contribute different non-zero indices from its own communication-set.
According to our observation, there are very few overlapping indices of the communication-set distribution of different nodes.
For example, training VGG16 on Cifar10 dataset using 16 GPUs with a compression ratio as 0.1\% for each node, the averaged compression ratio of synchronized residuals of all nodes is 1.55\%.
In this case, it is inefficient to use a sparse Allreduce~\cite{zhao2014kylix, renggli2018sparcml} for synchronization.
We utilize the Allgather operation, an operation in which the data contributed by each node is gathered at all nodes, to implement sparse Allreduce.
The compressed message representing communication-set of each node should include the information of indices and values of elements in communication-set.
When using threshold binary search selection, the length of each node's message is different.
As a result, the packaged message should also include an initial element, which indicates the length of the compressed elements.
Instead of using two Allgather operations for indices and values message separately, we package the indices and values into a single message to reduce latency.


After finishing the Allgather operation, each node collects $N$ compressed communication-sets from all the other nodes.
We add the compressed communication-sets to the corresponding weights in the local model after scaling with the learning rate.
It can be seen as an operation that adds a sparse array to a dense array, which has been fully-optimized in Level 1 function axpyi() of cuSparse library on GPU.

To analyze the potential performance gain of sparse synchronization, we adopt a performance model which is widely-used by~\cite{barnett1994interprocessor, mitra1995fast, thakur2005optimization} to analyze the cost in terms of latency and bandwidth used.
We assume that the time taken to send a message between any two nodes can be modeled as $\alpha + n\beta$,
where $\alpha$ is the latency (or startup time) per message, independent of message size, $\beta$ is the transfer time per byte, 
and $n$ is the number of bytes transferred. 
Generally, the node's network interface is assumed to be single ported, i.e. at most one message can be sent and one message can be received simultaneously. 
$M$ is the number of elements in residuals of the current layer.
$D$ is the compression ratio.
If we use \textit{threshold binary search} for communication-set selection, $D$ here should be the average compression ratio of all nodes.
In the case of reduction operations, 
we assume that $\gamma_2$ is the computational cost for performing the reduction operation for a message of size $M$, 
and $\gamma_1$ is the cost to decompress the collected sparse message of size $M$.

\begin{figure}[ht!]
\centering
\includegraphics[width=0.5\textwidth]{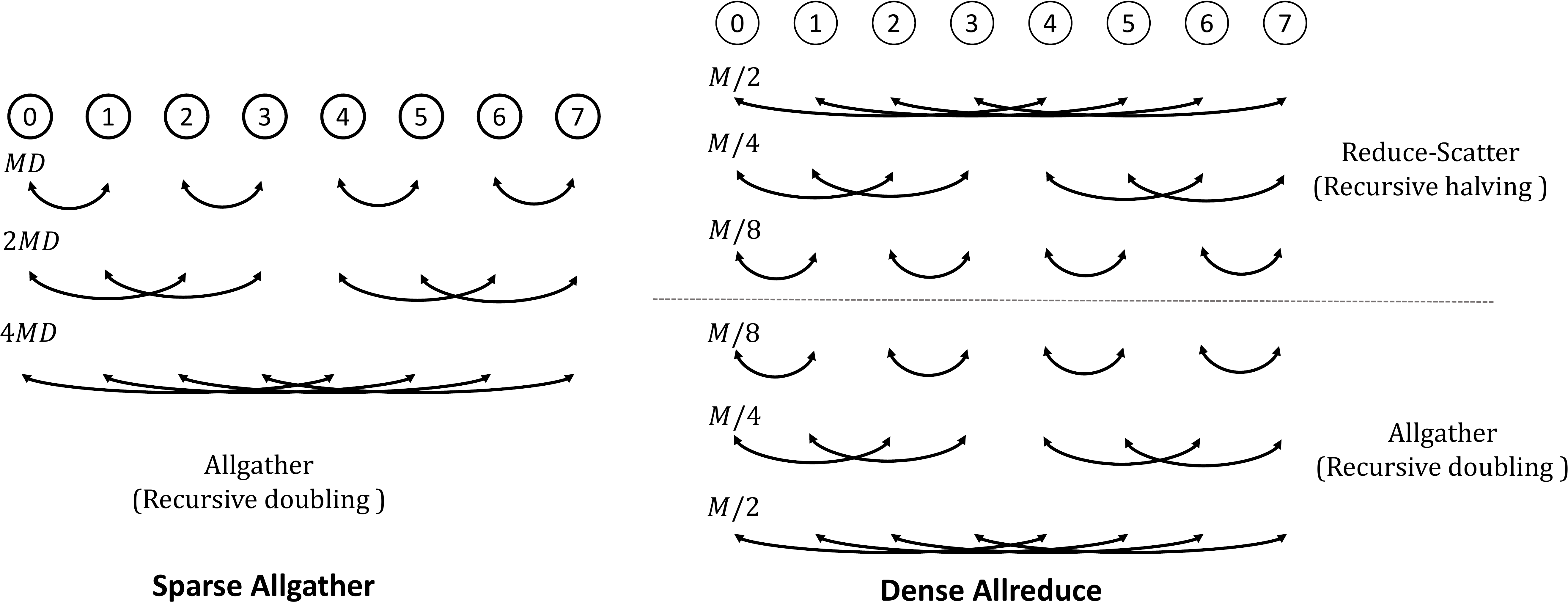}
\caption{Communication pattern of sparse synchronization with Allgather and dense synchronization with Allreduce.}
\label{fig:Allgatheralldreduce}
\end{figure}

Suppose that we use recursive doubling for Allgather and Rabenseifner's algorithm mentioned in \cite{thakur2005optimization} for Allreduce communication.
The cost of quantized sparse and dense synchronization is illustrated Equation (\ref{eq:sprs}) and Equation (\ref{eq:dns}), respectively.
The derivations are as follows:

The left part of Figure \ref{fig:Allgatheralldreduce} illustrates how sparse Allgather works by recursive doubling method.
In the first step, nodes that are a distance 1 apart exchange their compressed data, the size of which is $M \times D$.
In the second step, nodes that are a distance 2 apart exchange their own compressed data as well as the data they received in the previous step, which is $2M\times D$ in total.
In the third step, nodes that are a distance 4 apart exchange their own data as well the data they received in the previous two steps, which is $4M\times D$ in total.
In this way, for a power-of-two number of processes, all processes get all the data in lg$p$ steps. 
The amount of data exchanged by each node is $M\times  D$ in the first step, 2$M\times D$ in the second step, and so forth, up to $2^{lg(p)-1}M\times D$ in the last step. Therefore,
The time for message transfer taken by this algorithm is $T_{transfer}$ = $lg(p)\alpha + (p-1)M\times D \beta$.
After including decompressing overhead $\gamma$ and communication-set selection overhead $T_{select}$, the time for all-gather based synchronization should be 
$T_{Allgather}$ = $T_{select} + lg(p)\alpha + (p-1)M\times D \beta + p \gamma_1$.

As shown in the right part of Figure \ref{fig:Allgatheralldreduce}, the Rabenseifner's algorithm is adopted for Allreduce operation.
It does a reduce-scatter followed by an Allgather.
Reduce-scatter is a variant of reducing in which the results, instead of being stored at the root, are scattered among all $p$ nodes.
We use a recursive halving algorithm, which is analogous to the recursive doubling algorithm used for Allgather but in a reverse way.
In the first step, each node exchanges data with a node that is a distance $p/2$ away: 
Each process sends the data needed by all processes in the other half, which is of size $M/2$.
They also receive the data needed by all processes in its own half and performs the reduction operation on the received data. 
In the second step, each process exchanges data with a process that is a distance $p/4$ away.
This procedure continues recursively, halving the data communicated at each step, for a total of lg$p$ steps. 
After reduce-scatter, Allgather phase will have the the same bandwidth and latency requirements.
Therefore, the communication time taken by this algorithm is $T_{transfer}$ = $2lg(p)\alpha + 2\frac{p-1}{p}M\beta$.
The time taken by Allreduce is the sum of the times taken by reduce-scatter, Allgather and reduction operations.
The total time should be $T_{dense\_Allreduce}$ = $2lg(p)\alpha + 2\frac{p-1}{p}M\beta + \frac{p-1}{p}\gamma_2$.

\begin{equation}
\small
T_{sparse\_Allreduce} = T_{select} + \lg(p) \alpha + {(p-1)}{(MD)} \beta + p\gamma_1
\label{eq:sprs}
\end{equation}

\begin{equation}
\small
T_{dense\_Allreduce} = 2\lg(p) \alpha + 2\frac{p-1}{p} M\beta +  \frac{p-1}{p}\gamma_2
\label{eq:dns}
\end{equation}

As implicated by the performance model, two important conclusions about the current RGC algorithm can be drawn.
First, \textbf{the compression rate for the model is not equal to the compression rate for communication bandwidth.}
The bandwidth term of sparse synchronization is ($p-1$)$DM$$\beta$, which is proportional to the number of nodes $p$.
Even if the compression ratio $D$ is 0.1\% for all $p$ node, when $p$ is 128, the communication bandwidth for sparse synchronization will be 12.8\% of dense synchronization rather than 0.1\%.
Second,  \textbf{the overhead of decompression rather than communication will be a bottleneck when scaling RGC method to larger scale.}
The last term $p\gamma_1$ in Equation \ref{eq:sprs} indicates that the overhead of decompression in sparse Allreduce also increases linearly with the number of nodes $p$.
However, in Equation \ref{eq:dns}, reduction overhead of dense Allreduce almost does not increase with number of nodes.

\subsection{Overlapping Communication and Computation}
\label{sec:overlap}
It is necessary to improve data parallel efficiency by overlapping communication with computation through pipelining gradient Allreduce operations and backpropogation calculations.
Before updating aggregated gradients to weights, gradient clipping is usually adopted to avoid gradient explosion.
It rescales all of the gradients when the sum of their norms exceeds a threshold.
For RGC methods, the local clipping technique \cite{lin2017deep} is used to perform gradient clipping by a new threshold ($N^{-1/2}$ of the original one) locally before adding the current gradients to previous residuals.
The difference is that traditional data parallel does clipping after communication of all layers is completed, while the RGC algorithm needs to do clipping before communication.
In this case, we need to wait for the completion of the entire back-propagation to get gradients of all layers.
And then we do clipping on gradients and then perform compression for communication.
Local clipping introduces synchronization between computing and communication and thus eliminates the overlapping of communication and computation.

\begin{figure}[ht!]
\centering
\includegraphics[width=0.5\textwidth]{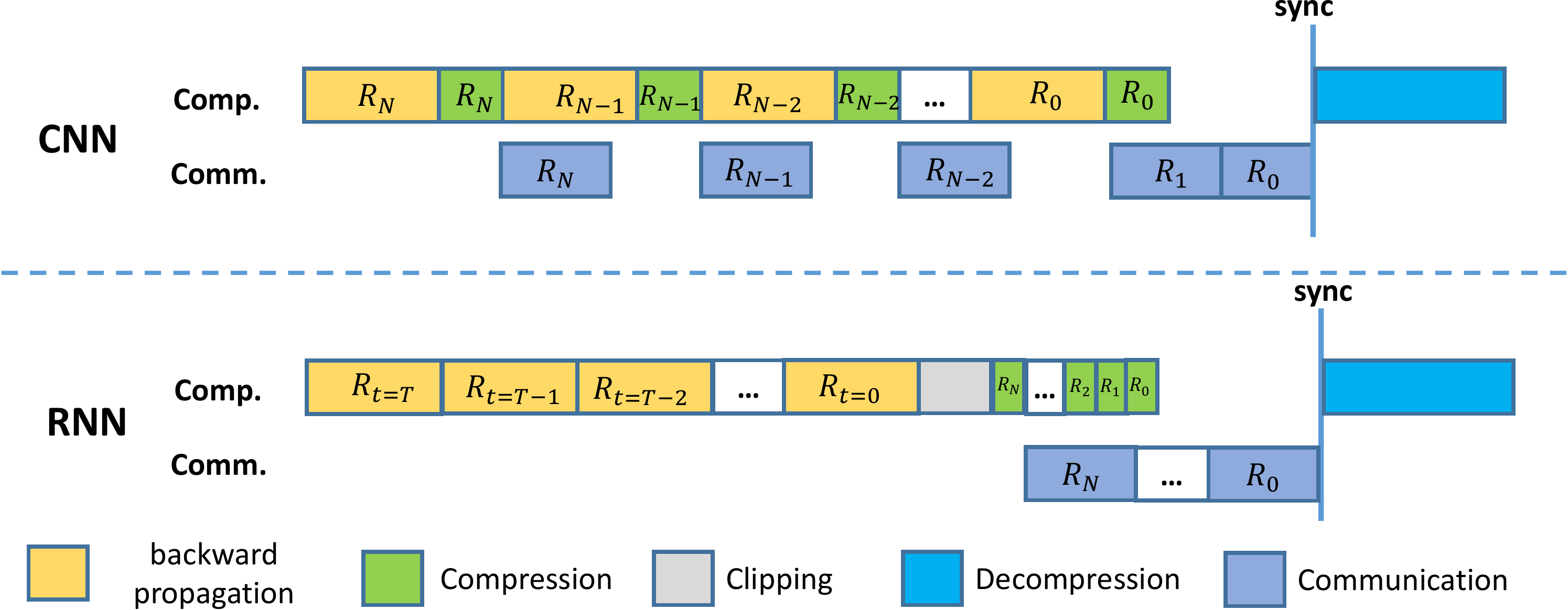}
\caption{CNNs and RNNs adopt different schemes to overlap communication with computation.}
\label{fig:overlap}
\end{figure}

As shown in Figure \ref{fig:overlap}, 
RedSync has abandoned gradient clipping for CNNs, which seldom have gradient exploration problem in order to explore the potential overlapping.
As for RNNs, gradients are achieved after backpropagation of all time steps using Back Propagation Through Time (BPTT).
When backpropagation of the last time step is completed, gradients of all layers are used to conduct local gradient clipping.
In this case, the communication time can only overlap with the compression (selection) operation.

\subsection{Other techniques}
RedSync supports a set of algorithmic improvements proposed by DGC\cite{lin2017deep} to avoid convergence problem.
For momentum SGD and Nesterov momentum SGD optimizers, the momentum masking and momentum correction schemes of DGC are implemented by simply modifying workflow of Algorithm \ref{algo:rgcframework}.
A warm-up training, by exponentially decreasing the compression ratio in the first few epochs, is generally adopted to accelerate convergence.
For example, it is recommended to decrease the compression ratio in the warm-up period as follows: 25\%, 6.25\%, 1.5625\%, 0.4\%, 0.1\%.
However, according to our performance model of communication, such an approach is inefficient on a large scale.
For example, synchronization with a compression ratio as 1.5625\% requires 100\% bandwidth of dense Allreduce for quantized RedSync on 64 GPUs.
Instead of adopting a high-compression-ratio RGC method of warm-up training, we use original SGD optimizer synchronized by dense Allreduce in first few epochs if necessary.


%

\section{Experimental Results}
\subsection{Setups}
We tested the accuracy, speed of convergence and scalability of RedSync on two different multi-GPU systems, including a world's top GPU supercomputer (Piz Daint) and a multi-GPU server (Muradin).

\textbf{Muradin} is a server with eight GPUs in the same node.
It is equipped with one Intel(R) Xeon(R) CPU E5-2640 v4 and 8 TITAN Vs, which is connected to the CPU through PCI-E 3.0.

\textbf{Piz Daint} is a GPU supercomputer.
Each node of it includes two Intel Xeon E5-2690v3 CPUs and one NVIDIA Tesla P100 GPUs.
In total, there are 5320 nodes connected by Aries interconnect with Dragonfly topology.

We used pytorch v0.4 to conduct DNN training on a single GPU.
In terms of the communication library, an MPI wrapper upon pytorch called horovod \cite{sergeev2017meet}, is used to provide collective communication operations.
The CUDA version is 9.1 on Muradin and 8.0 on Piz Daint.
Horovod was compiled with OpenMPI v3.1 with cuda-aware supported on both systems.
%

Two major types of mainstream deep learning applications are used in the experiments.
For \textbf{Image Classification} tasks,
we studied ResNet44  and VGG16 on Cifar10\cite{krizhevsky2009learning}, AlexNet, VGG16 and ResNet-50 on ImageNet\cite{deng2009imagenet}.
For all CNNs, we used Nesterov's momentum SGD as the optimizer.
RGC methods used the same learning rate strategies as SGD.
The warm-up technique was applied to the first 5 epochs of ResNet50 and VGG16 for both SGD and RGC.
For \textbf{Language Modeling}  tasks, we picked two datasets for evaluation.
The Penn Treebank corpus (PTB) dataset \cite{marcus1993building} consists of 923,000 training, 73,000 validation and 82,000 test words. 
The WikiText language modeling dataset is a collection of over 100 million tokens extracted from the set of verified Good and Featured articles on Wikipedia \cite{merity2016pointer}.
It consists 2,088,628 training, 217,646 and 245,569 test words.
We adopted a 2-layer LSTM language model architecture with 1500 hidden units per layer \cite{press2016using} to evaluate both datasets. 
We tied the weights of encoder and decoder and use vanilla SGD with gradient clipping.
Learning rate decays when no improvement has been made in validation loss.
For RedSync, we used trimmed top-0.1\% selection for convolutional layers larger than 128 KB and used threshold binary search top-0.1\% selection for hidden layers and the softmax layer of LSTM.

\subsection{Evaluation of Accuracy and Convergence Speed}
\begin{table}[ht!]
\small
\caption{Accuracy Results for Various DNNs. }
\centering
\begin{tabular}{|c||c|c|c|c|c|c|c|}
\hline
\bf  \multirow{ 2}{*}{Dataset}  & \bf  \multirow{ 2}{*}{DNN}  & \bf  \multirow{ 2}{*}{Size}   & \bf  \multirow{ 2}{*}{Gflops}    & \multicolumn{3}{c|}{\bf  Accuracy} \\
\cline{5-7}
 & \bf  & \bf   & \bf  & \bf SGD & \bf RGC & \bf  RGC+ASQ \\
\hline
\hline
\multirow{ 2}{*}{Cifar10} & ResNet44 & 2.65 &  0.20 & 7.48\% & \textbf{7.17\%} & 7.87\% \\
\cline{2-7}
& VGG16 & 59 & 0.31 & 8.31\% &  8.45\% & \textbf{ 8.13\%} \\
\hline
\multirow{ 2}{*}{ImageNet} & AlexNet & 233  & 0.72 & \textbf{44.73\%} & 44.91\% & 44.80\% \\
\cline{2-7}
& ResNet50 & 103 & 8.22 & 24.07\% &  23.98\%  & \textbf{23.85\%} \\
\cline{2-7}
& VGG16 & 528 & 15.5 & 29.5\% &  \textbf{29.1\%}   & 29.3\%\\
\hline
PTB & LSTM & 204 & 2.52  & 75.86 &  {75.14} & \textbf{74.69} \\
\hline
Wiki2 & LSTM & 344 & 2.52  & 88.23 &  88.01 & \textbf{87.84} \\
\hline
\end{tabular}
\label{tab:acc1}
\end{table}

As shown in Table \ref{tab:acc1}, we examined the accuracy of both RGC and our proposed quantization version RGC+ASQ on Muradin.
The results of RGC and RGC+ASQ are compared with a classical data parallel implementation using SGD.
Size indicates the model size in Megabyte.
GFlop shows Giga Floating-Point Operations required for a forward pass using a single input sample.
Accuracy of CNNs was measured as top-1 validation errors, and accuracy of LSTMs is measured as perplexity on validating dataset.
Results on Cifar10 were achieved using 4 nodes (batch-size\footnote{The batch-size here is for a single node.} = 64).
Results on ImageNet were achieved using 6 nodes (batch-size = 32).
Results of LSTM were achieved using 4 nodes (batch-size = 5).

As implicated by Table \ref{tab:acc1}, 
the accuracy of the models obtained by the RGC and RGC+ASQ methods using RedSync is similar to that obtained by the SGD method using classical data parallel, with a difference of no more than 1\%.
In only one case (ImageNet-AlexNet), SGD achieves the best accuracy results, and in the other six cases, RGC and RGC+ASQ are even slightly better than SGD.
In three cases (Cifar10-VGG16, ImageNet-ResNet50, Wiki2-LSTM), RGC+ASQ is slightly better than RGC, which indicates that the ASQ quantization method proposed in this chapter is very reliable and has no impact on training accuracy.

\begin{table}[ht!]
\small
\caption{Accuracy of RGC and SGD Methods Under Different Batch Size.}
\centering
\begin{tabular}{|c||c|c|c|c|c|c|c|}
\hline
& Batch Size & 128 & 256 & 512 & 1024 & 2048\\
\hline
\hline
\multirow{3}{*}{ResNet44} & SGD & 7.09\% & 7.48\% & 8.18\% & \textbf{10.02\%} & 10.84\% \\
\cline{2-7}
& RGC & \textbf{6.40\%} & \textbf{7.17\%}  & \textbf{7.47\%} & 10.13\% & 10.87\% \\
\cline{2-7}
& RGC+ASQ & 7.06\% & 7.87\% & 7.62\% & 11.86\% & \textbf{10.83\%} \\
\hline
\hline
\multirow{3}{*}{VGG16} & SGD & 7.74\% & 8.31\% & \textbf{9.06\%}	 & \textbf{9.49\%} & 10.09\% \\
\cline{2-7}
& RGC & \textbf{7.43\%} & 8.45\% & 9.31\% & 9.90\% & 11.12\% \\
\cline{2-7}
& RGC+ASQ & 8.17\% & \textbf{8.13\%} & 9.09\% & 9.97\% & \textbf{9.81\%} \\
\hline
\end{tabular}
\label{tab:acc2}
\end{table}

\begin{figure*}[ht!]
\centering
\includegraphics[width=0.75\textwidth]{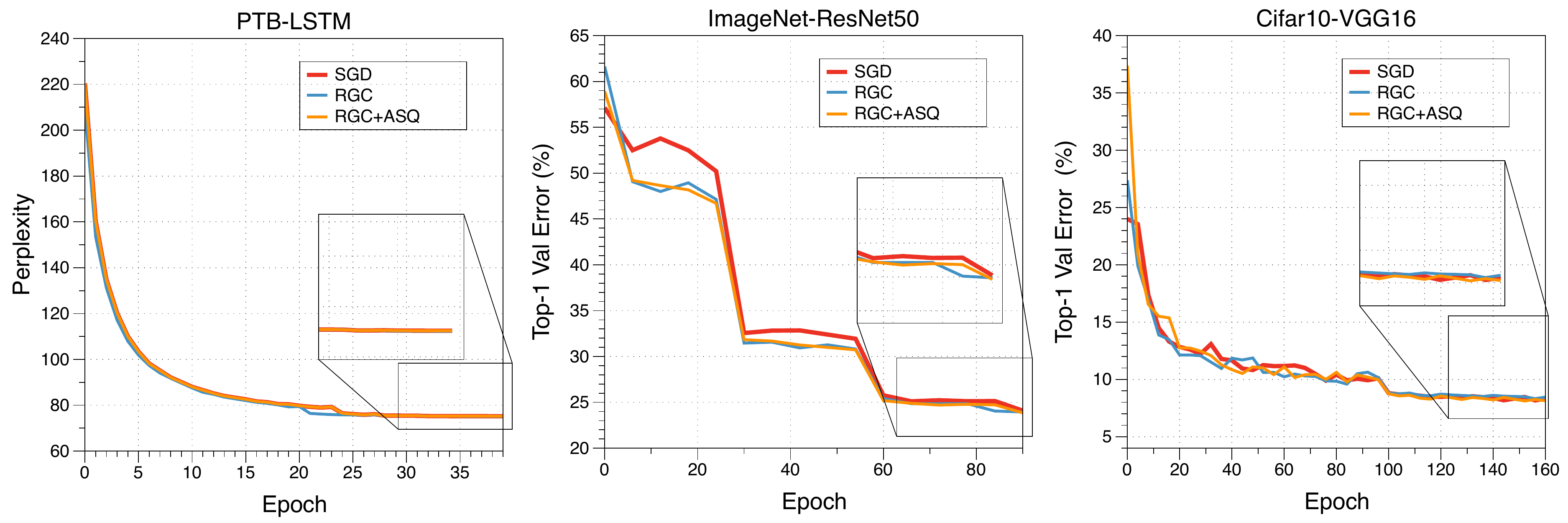}
 \caption{
Speed of Convergence using RedSync compared with classical data parallel implementation.
}
\label{fig:acccurv}
\end{figure*}

Figure \ref{fig:acccurv} shows the speed of convergence of RGC and RGC+ASQ implemented with RedSync on three typical test cases compared with SGD implemented with the classical data parallel.
The left figure shows top-1 validation accuracy vs the number of epochs of training VGG16 on Cifar10 (batch-size = 64). 
The center figure shows top-1 validation accuracy vs the number of epochs of training ResNet50 on ImageNet (batch-size = 32). 
The right figure shows Perplexity vs the number of epochs of training LSTM on PTB (batch-size = 5).

As shown in Table \ref{tab:acc2}, we also tested the sensitivity of the RGC and RGC+ASQ methods to large training data batch size.
when increasing the batch size to 2048, RedSync got no loss of accuracy compared to the original SGD.

\subsection{Evaluation of Scalability}
\begin{figure*}[ht!]
\centering
\includegraphics[width=0.78\textwidth]{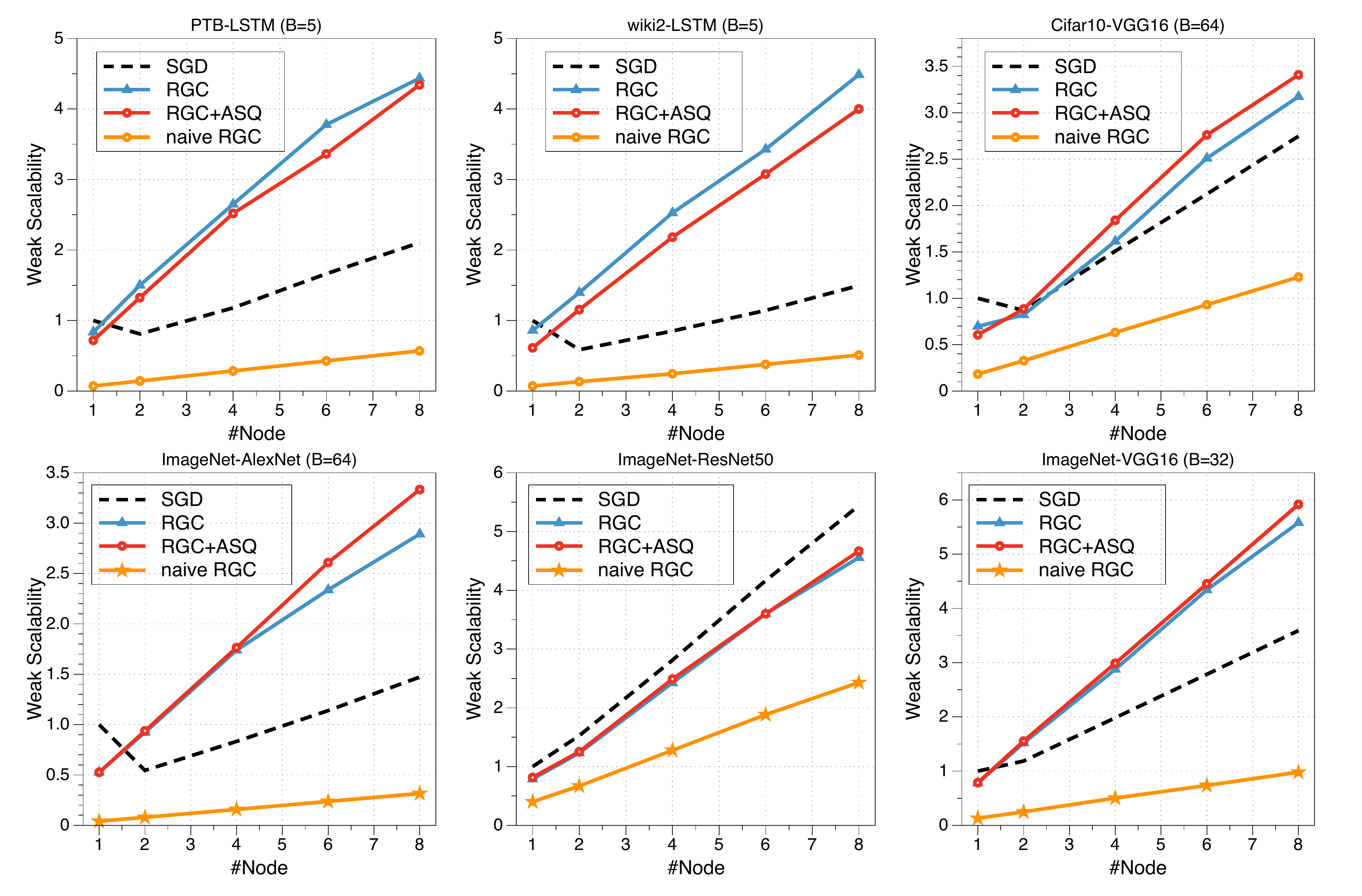}
\caption{Scalability of RedSync for CNNs and RNNs training using Muradin.}
\label{fig:muradin}
\end{figure*}

\begin{figure*}[ht!]
\centering
\includegraphics[width=0.78\textwidth]{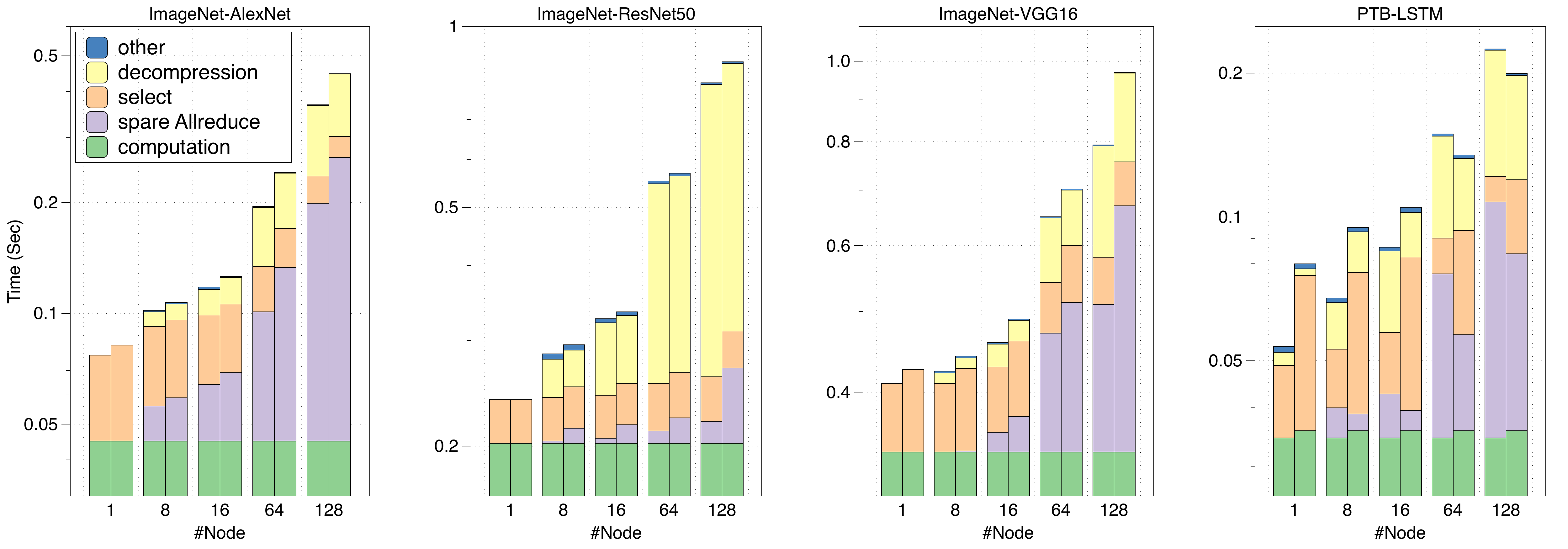}
\caption{Decomposition of time cost of RedSync on Piz Daint. For each two-column group, the left one is for RGC+ASQ and the right one is for RGC. Y-axis is presented as 10 average iteration time.}
\label{fig:percent}
\end{figure*}

\begin{figure*}[ht!]
\centering
\includegraphics[width=0.80\textwidth]{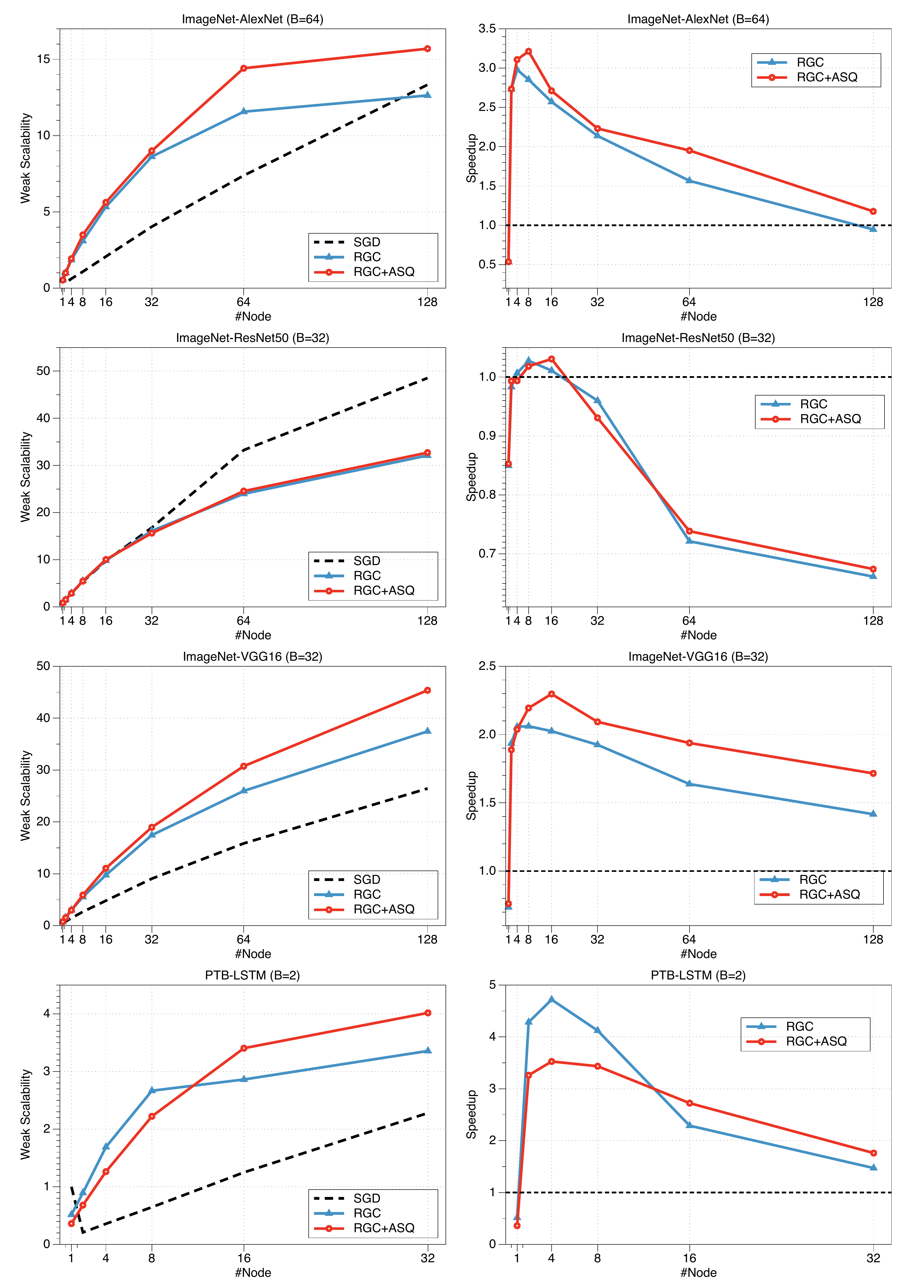}
\caption{Left: Scalability of RedSync for four DNNs on Piz Daint. Right: Speedup of RedSync compared with a classical data parallel implementation.}
\label{fig:daint}
\end{figure*}

To evaluate the scalability of RedSync on different scales, we compare RGC and RGC+ASQ implemented by RedSync with SGD implemented by the data parallel scheme provided by horovod.
The performance was measured by averaging training time of 1000 iterations.
Figure \ref{fig:muradin} shows the performance of RedSync on Muradin with six test cases.
Figure \ref{fig:daint}  illustrates the scalability of RedSync on Piz Daint with four test cases.
In order to find time-consuming parts, we also illustrate the cost of different parts in RedSync in Figure \ref{fig:percent} when scaling it to 128 GPUs on Piz Daint.
Our observations are summarized as follows.

Our proposed parallel-friendly selection algorithms for gradient sparsification are critical for improving the overall performance of RGC-based system.
In Figure \ref{fig:muradin}, we added a naive RGC implementation, which uses radixSelect to select top 0.1\% elements as communication-set rather than our proposed methods.
Since the compression time is too long, the performance of the naive RGC is even much slower than the original data parallel SGD version.
After adopting our top-0.1\% selection algorithms, the RGC and RGC+ASQ systems are now able to run faster than the SGD version.

Our proposed RGC+ASQ method is better than RGC-only method in most cases.
RGC+ASQ always achieves better performance than RGC for CNNs.
However, for LSTM training on a small scale, RGC+ASQ achieves worse performance than ASQ.
The variance of communication and computational overhead accounts for such a phenomenon.
CNN adopts trimmed top-0.1\% as the communication-set selection method and its quantized version (RGC+ASQ) has a similar computation cost.
As shown in Figure \ref{fig:percent}, no significant difference of \textit{selection} cost between RCG and RGC+ASQ in CNN training.
Therefore, the reducing of communication cost by ASQ improves the system's overall performance.
As for selection algorithm of LSTMs, RGC uses sampled threshold binary search selection as communication-set selection, but RGC+ASQ uses threshold binary search.
As we mentioned, the sampled selection is much faster.
Therefore, on small-scale, RGC has better scalability than RGC+ASQ due to less selection overhead.
When scaling to more than 16 GPUs, benefit from the reduction of communication bandwidth compensates for the cost of the communication-set selection.

RedSync is suitable for training DNNs of high communication to computation ratio.
In the past, these DNNs have been considered as not suitable for classical data parallel.
As shown in the Figure \ref{fig:daint}, for VGG16, AlexNet, and LSTM, although the performance of RedSync using RGC and RGC+ASQ on a single GPU are not as good as the data parallel due to compression and decompression overhead, 
RedSync can achieve significant speedup on more than 2 GPUs.
An exception is ResNet50.
in most of the cases, RedSync brings no performance gain for it both on Piz Daint and Muradin.
As implicated in Table \ref{tab:acc1}, the ratio of computation to the communication of ResNet50 is the highest in the DNNs we investigated.
On a large scale, most of the time during ResNet50 training with RedSync is wasted on the decompression phase, as shown in Figure \ref{fig:percent}, which overshadows the benefit of communication bandwidth reduction.

RedSync is suitable for the medium parallel scale.
As shown in the right part of Figure \ref{fig:daint}, 
for the AlexNet network, RedSync is strong (brings the most significant performance improvement) on 4-16 nodes, bringing around 3x speedup.
For the ResNet50 network, RedSync is strong on 4-16 nodes, bringing around 1.2x speedup.
For the VGG16 network, RedSync is strong on 4-64 nodes, bringing around 2x speedup.
For LSTM networks, RedSync is strong on 2-16 nodes, bringing around 3x speedup.
Since both communication bandwidth requirements and decompression overhead increase linearly with the number of GPUs in use, RedSync is relatively weak on a larger scale.
This phenomenon just confirms the communication performance model proposed in Section \ref{sec:sparseAllreduceComm}.

%


\section{Related Works}
As shown in Table \ref{tab:relatedworks}, the related works to reduce communication cost in data parallel training of DNNs can be divided into two categories: quantification (Qunat.) and sparsification (Spars.).
The compression ratio (Ratio in the table) is for a single node.
Impl. in the table Indicates whether the method has been implemented in a real distributed environment.

\begin{table}[ht!]
\small
\centering
\caption{Related Works.}
\begin{tabular}{|c||c|c|c|c|c|c|c|}
\hline
Name & Spars. & Quant. & 1/Ratio & Sync. & Impl. & Scale\\
\hline
\hline
1-bit SGD\cite{seide20141} & $\times$  & $\surd$ & 32x & PS & $\surd$ & 40 GPU\\
\hline
QSGD\cite{alistarh2017qsgd} & $\times$  & $\surd$ & 4-6.8x & PS & $\surd$ & 16 GPU\\
\hline
Strom\cite{strom2015scalable} & $\surd$ & $\surd$ & 800x & PS & $\surd$ & 80 GPU\\
\hline
AdaComp\cite{chen2018adacomp}  & $\surd$ &  $\times$  & 40-200x & PS & $\times$ & -\\
\hline
TernGrad\cite{wen2017terngrad} &  $\times$  & $\surd$ & 16x & PS & $\times$ & - \\
\hline
DGC\cite{lin2017deep} & $\surd$ &  $\times$  & 1000x & AllRed. & $\times$ & -\\
\hline
SparCML\cite{renggli2018sparcml} & $\surd$ & $\surd$  & 256x & AllRed. & $\surd$ & 128 GPU\\
\hline
\end{tabular}
\label{tab:relatedworks}
\end{table}

In terms of synchronization scheme (Sync.), DGC and SparCML use Allreduce and the other methods adopt Parameter Server (PS).
Allreduce is more suitable for HPC platforms, while PS can hardly benefit from efficient Allreduce routines designed for HPC network hardware.
For Allreduce systems, according to analysis in Section \ref{sec:sparseAllreduceComm}, it should be distinguished from the compression ratio of communication bandwidth.
Similar for PS systems, the local compression ratio only reflects the bandwidth reduction of pushing gradients of works to the servers, while the bandwidth requirement of pulling gradients from the servers to works also increases linearly with the number of nodes.
RedSync is the best in terms of scale compared with the other distributed implementations.
SparCML also scales DNN training to 128 GPUs, 
but it adopts a "fast randomized top-$k$ algorithm", which is not equal to the top-$k$ selection.
Its communication-set misses some important gradient elements and the system will suffer from convergence problem.
Both Storm and SparCML combine quantization technique with sparsification, but in a less efficient way than our proposed ASQ method.
As we mentioned, Storm's method is less efficient than ASQ, while SparCML quantizes the value elements of communication-set using 4-bit precision.

In terms of accuracy, DGC, TernGrad, and AdaComp presented comprehensive evaluation results on classical CNNs and RNNs.
The other methods are either tested on some specified DNN models or have relative large accuracy loss compared with SGD. 
RedSync has been comprehensively evaluated using most of the test cases used in DGC.
It is unfair to compare the absolute speedup achieved by each proposed method,
because the achievable benefit is dependent on multiple factors of experimental settings, including the type of DNN model, the bandwidth of network hardware and the speed of computing hardware.
The Aries interconnect of Piz Daint used in our experiments is the most high-quality one compared with others, which means RedSync is able to achieve more performance gain on low-quality network hardware.

\section{Conclusion}
This paper proposes an innovative data parallel DNN training design called RedSync that compressing transmitting data by gradient sparsification and quantization.
Residual Gradient Compression (RGC) method is used for sparsification.
RedSync solved two major obstacles to implement RGC on multi-GPU systems: the high overhead of communication-set selection on GPU and lack of support for collective communication scheme for sparse data structures.
Based on RGC, Alternating Signs Quantization (ASQ) method is first proposed, which further reduces half of the communication bandwidth requirement and introduces no extra overhead.
The performance and accuracy of RedSync are evaluated on two typical GPU platforms, including a supercomputer system and a multi-GPU server.
For AlexNet, VGG16, and LSTM, RedSync brings significant speedup for large-scale DNN training.